\documentclass[a4paper, 12pt]{article} % Font size (can be 10pt, 11pt or 12pt) and paper size (remove a4paper for US letter paper)
\usepackage[right=1in,left=1in,top=0.93in,bottom=0.94in]{geometry}
\usepackage{graphicx} 
\usepackage{braket}
\usepackage{mathrsfs}
\usepackage[T1]{fontenc} % Required for accented characters
\usepackage{amsmath}
\usepackage{pxfonts}
\usepackage[T1]{fontenc}
\usepackage{amssymb}
\usepackage{natbib}
\usepackage{epstopdf}
\usepackage{setspace}
\usepackage{enumitem}
\bibpunct{(}{)}{,}{a}{,}{,}
\usepackage{sectsty}
\usepackage{wrapfig} % Allows in-line images
\sectionfont{\large}
\usepackage{tcolorbox}
    \usepackage{pgfplots}

\usepackage{tikz}   %TikZ is required for this to work.  Make sure this exists before the next line
\usepackage{tikz-3dplot} %requires 3dplot.sty to be in same directory, or in your LaTeX installation
\newcommand{\qed}{\nobreak \ifvmode \relax \else
      \ifdim\lastskip<1.5em \hskip-\lastskip
      \hskip1.5em plus0em minus0.5em \fi \nobreak
      \vrule height0.75em width0.5em depth0.25em\fi}

\usepackage{bbm}
\usepackage{MnSymbol}
\usepackage[all]{xy}

\usepackage{xcolor,colortbl,array,amssymb}

\usepackage{centernot}

\usepackage{epigraph}
\makeatletter

\newlength{\bibitemsep}
\newlength{\bibparskip}
\let\oldthebibliography\thebibliography
\renewcommand\thebibliography[1]{%
  \oldthebibliography{#1}%
  \setlength{\parskip}{\bibitemsep}%
  \setlength{\itemsep}{\bibparskip}%
}

\newcommand{\x}[1]{{#1}}

\title{\textbf{The Simplicity of Physical Laws}}

\author{Eddy Keming Chen\thanks{Department of Philosophy,  University of California, San Diego, 9500 Gilman Dr, La Jolla, CA 92093-0119. Website: www.eddykemingchen.net. Email: eddykemingchen@ucsd.edu  }
}

\date{\textit{No\^us}, final version of \today}

\begin{document}

\maketitle % Print the title section

%----------------------------------------------------------------------------------------
%	ABSTRACT AND KEYWORDS
%----------------------------------------------------------------------------------------

%\epigraph{Simplicity is not a simple thing.}{Charlie Chaplin}

%\newpage

\begin{abstract}

Physical laws are strikingly simple, yet there is no \textit{a priori} reason for them to be so. I propose that nomic realists---Humeans and non-Humeans---should recognize simplicity as a fundamental epistemic guide for discovering and evaluating candidate physical laws. This proposal helps resolve several longstanding problems of nomic realism and simplicity. A key consequence is that the presumed epistemic advantage of Humeanism over non-Humeanism dissolves, undermining a prominent epistemological argument for Humeanism. Moreover, simplicity is shown to be more connected to lawhood than to mere truth.  

\end{abstract} 

%\hspace*{3,6mm}\textit{Keywords: empirical equivalence, induction, underdetermination, determinism, strong determinism, Humeanism, non-Humeanism, laws of nature, ratbag idealism, Bayesianism, comparative probability, expert principle, theoretical virtues, IBE, nested theories, vagueness }   % Keywords

\begingroup
\singlespacing
\tableofcontents
\endgroup

%\vspace{30pt} % Some vertical space between the abstract and first section

%----------------------------------------------------------------------------------------
%	ESSAY BODY
%----------------------------------------------------------------------------------------
\nocite{HumeTreatise, humeenquiry}

%Is PNS required for induction or nomic realism? I offer this framework as one possible and natural ground for justification. 

\section{Introduction}

Physical laws are strikingly simple, yet there is no \textit{a priori} reason for them to be so.  My goal in this paper is to articulate and clarify the view that simplicity is a fundamental epistemic guide for discovering and evaluating the fundamental laws of physics. 

Many physicists and philosophers are realists about physical laws. Call realism about physical laws \textit{nomic realism}. It consists of two components: first, that physical laws are objective and mind-independent; second,  that we have epistemic access to physical laws.\footnote{See \cite{peebles2024physicists} for a recent example of a physicist's version of nomic realism.} 

Nomic realism appears to face an epistemic gap: if physical laws are objective and mind-independent, it remains puzzling how we have epistemic access to them, given that they do not directly follow from our observations. This gap exemplifies a broader challenge in scientific realism \citep{sep-scientific-realism} regarding the justification of theoretical statements. 

To address this epistemic gap, nomic realists appeal to super-empirical theoretical virtues, a familiar example of which is simplicity. Simplicity allows us to  eliminate a vast range of empirically equivalent theories that posit unnecessarily complex laws. Combined with other theoretical virtues, simplicity may even help narrow down the choice to a unique theory given the totality of evidence. However, simplicity itself presents several philosophical difficulties \citep{sep-simplicity, iep-simplicity}:

\begin{enumerate}
  \item The problem of coherence: Naive applications of simplicity can lead to probabilistic incoherence (which is sometimes called the problem of nested theories, or the problem of conjunctive explanations).  
  \item The problem of justification: There is no plausible epistemic justification for  simplicity as a guide to truth. 
  \item The problem of precision: There is no universally accepted precise standard for measuring simplicity. 
\end{enumerate}
These challenges are not unique to simplicity; similar questions arise for other super-empirical virtues, such as unification and informativeness. We face related problems whenever we use super-empirical virtues to guide our theory choice.  What one says about simplicity should also apply to other epistemic guides that nomic realism employs.  Given its widespread interest and methodological importance, this paper focuses on simplicity. 

I develop a framework for understanding simplicity as a fundamental epistemic guide to physical laws. By analyzing the commitments of nomic realism, we uncover a straightforward resolution to the problem of coherence. Reflection on the applications of simplicity provides strong reasons to take it as a \textit{fundamental} principle requiring no further epistemic justification. Moreover, examining the diversity of cases where simplicity applies reveals the necessity of treating it as a \textit{vague} principle that acquires precise forms only in specific contexts.

Is this framework redundant for Humeans? No. It is sometimes believed that, on the best-system account of laws (BSA), we get the simplicity of physical laws for free, because laws are defined to be simple and informative summaries of the mosaic. However, that is a mistake, since we are not directly given the mosaic. The principle of simplicity must be added to BSA as an epistemic norm, guiding our expectations about the best system, even when we do not have direct epistemic access to the mosaic. As we shall see, since both Humeanism and non-Humeanism need an independent epistemic principle concerning the simplicity of physical laws, they are on a par regarding the empirical discovery of laws. If Humeans can adopt the epistemic guide without issue, non-Humeans can as well. As a result, the frequently cited epistemic advantage of the former over the latter vanishes, undermining a key epistemological argument for Humeanism \citep{earman2005contact}. (See \cite[\S8]{hildebrand22} for a similar perspective; see also \cite[\S4.1]{chenandgoldstein}.) Nevertheless, the simplicity postulate may target different aspects: in non-Humeanism, it serves as an epistemic guide for selecting laws, whereas in Humeanism, it ultimately guides which types of mosaics should be taken seriously. 

Recent work in the foundations of physics and the metaphysics of laws offers new case studies that highlight the need for a  systematic treatment of the methodological principles underpinning nomic realism.  This paper presents such a framework. This may not be the only approach possible, but it has a number of features attractive to nomic realists. For one thing,  simplicity is recognized as an important theoretical virtue in scientific practice \citep{schindler2022theoretical}, and it is one principle that nomic realists often explicitly endorse. Its theoretical benefits, as I hope to show, justify the cost of the posit. Although the discussion here is not meant to convince nomic anti-realists, they may still find it useful for understanding a position they ultimately reject. 

Here is the plan. First, I clarify nomic realism and its  metaphysical and epistemological commitments, illustrating  the epistemic gap  with three algorithms for generating empirical equivalent theories. I introduce the principle of nomic simplicity as a tie-breaker and show how it resolves the problem of coherence. 
Next, I examine five further applications of the principle---induction, symmetries, dynamics,  determinism, and explanation---arguing that simplicity plays a more fundamental role than many widely accepted methodological assumptions. 
Finally, I suggest that the principle is best understood as a fundamental yet vague epistemic guide and clarify why Humeanism holds no  epistemic advantage over non-Humeanism in our access to physical laws. 

%\begin{enumerate}
%  \item To say that simplicity is an epistemic guide is to say that it is truth-conducive, i.e. it is a general guide to truths in the world. 
%  \item We cannot appeal to simplicity unless we can show how it is to be reduced to evidential support. 
%  \item If simplicity is too vague, then we cannot appeal to it. 
%  \item Regarding simplicity as an epistemic guide is a monopoly of the Humean best-system views, and non-Humeans worth their salt cannot and should not lay claim on simplicity.
%  \item Humeans have a principled explanation for why laws are simple while non-Humeans cannot. And this shows Humeans have an epistemic advantage over non-Humeans. 
%\end{enumerate}

\section{Nomic Realism}

Nomic realism appears to face an epistemic gap.  It arises from the difficulty of reconciling two central commitments of \textit{nomic realism}: 

\begin{description}
  \item[Metaphysical Realism:] Physical laws are objective and mind-independent; more precisely, the propositions that express physical laws correspond to objective and mind-independent facts about the world.\footnote{A weaker version of metaphysical realism allows that laws are not entirely mind-dependent. This accommodates more pragmatic versions of the Humean best-system accounts (e.g. \cite{loewer2007laws, cohen2009better, hicks2016dynamic, dorst2019towards, jaag2020making}, and the volume edited by \cite{hicks2023humean}), in which the mosaic still partially determines the best system. The arguments below should apply with suitable modifications.}
\end{description}

\begin{description}
  \item[Epistemic Realism:] We have epistemic access to physical laws; more precisely, we can be epistemically justified in believing which propositions express physical laws, given the evidence that we will in fact obtain.\footnote{The terminology follows \cite{earman2005contact}.  I have added the clause `given the evidence that we will in fact obtain,'  making my version of epistemic realism logically stronger than theirs, which considers all possible evidence we might obtain. I return to this distinction in \S5.3. }  
\end{description}
Nomic realists seek to endorse both theses, but the challenge lies in explaining how. The issue is an instance of a broader epistemological problem: how we can be justified in believing propositions that extend beyond the logical closure of empirical evidence? It is closely related to issues about the rationality of induction and scientific explanation, which  will be explored in \S4. However, before addressing these broader concerns, it is necessary to understand how the epistemic gap manifests in specific cases.   For concreteness, let us examine a Humean account and a non-Humean account, both of which aspire to satisfy nomic realism. \x{The epistemic gap, shared by many realist accounts of physical laws, can be illustrated with the following examples.}

\subsection{Two Accounts}

First, consider the Humean best-system account of \cite{LewisCounterfactuals, LewisNWTU, lewis1986philosophical}, with some  modifications: 
\begin{description}
  \item[Best System Account (BSA)]   Fundamental laws of nature are the axioms of the best system that summarizes the mosaic and optimally balances simplicity, informativeness, fit, and degree of naturalness of the properties referred to. The mosaic (spacetime and its material contents) contains only local matters of particular fact, and the mosaic is the complete collection of fundamental facts. The best system supervenes on the mosaic.\footnote{A key difference between this version and Lewis's \citep{LewisCounterfactuals, LewisNWTU, lewis1986philosophical} is that his original formulation requires fundamental laws to be regularities, whereas the version here drops the requirement. Another difference is the replacement of perfect naturalness with degree of naturalness. See \cite[sect.2.3]{chen2018NV} for further discussion. On Humeanism, the mosaic is traditionally required to be about local matters of particular fact. There are other Humean accounts of laws; see \cite{roberts2008law} for an example. }
  \end{description}
  BSA satisfies metaphysical realism, even though its laws are not metaphysically fundamental. Given a particular mosaic (spacetime manifold with material contents), there is a unique best system that is objectively best.\footnote{For the sake of the argument, for now I set aside the worry of ``ratbag idealism'' and  grant Lewis's assumption that nature is kind to us \cite[p.479]{LewisHSD}. Even if a revised version of BSA satisfies only the weaker version of metaphysical realism for which laws are not entirely mind-dependent, it does not automatically secure epistemic realism, as we shall see.}  

Next, consider a recent non-Humean account, which holds that laws govern and exist over and above the material contents  \citep{chenandgoldstein}: 
  \begin{description}
  \item[Minimal Primitivism (MinP)] Fundamental laws of nature are certain primitive facts about the world. There is no restriction on the form of the fundamental laws.  They govern the behavior of material objects by constraining the physical possibilities.  
\end{description}
MinP satisfies metaphysical realism, because the primitive facts corresponding to laws are objective and mind-independent.  It is minimal in that it does not restrict the form of laws.\footnote{In this paper, I shall use ``fundamental laws,'' ``fundamental physical laws,'' ``physical laws,'' and ``laws'' interchangeably.} MinP is compatible with fundamental laws taking on the form of  boundary conditions, least action principles, and global spacetime constraints.\footnote{See \cite{adlam2022laws} and \cite{meacham2023nomic} for related views, and \cite{hildebrand2020non} for an overview of non-Humeanism. \x{The arguments below, with suitable modifications, apply to other versions of non-Humeanism, including the Powers Best-System Account (Powers BSA) developed by \cite{demarest2017powerful} and \cite{kimpton2017humean}. There is also an epistemic gap between our evidence and the best systematization of the exact power distributions in the actual world and other possible worlds. See \cite{SchwarzKnowing2023} for further discussion. Defenders of Powers BSA can adopt a version of PNS to guide their epistemic expectations about the likely power distributions.  } }  \cite{chenandgoldstein} also introduce an epistemic principle called ``Epistemic Guides,'' which we will discuss in \S5.3.

\subsection{The Epistemic Gap}

Do BSA and MinP support epistemic realism? Their metaphysical commitments alone do not guarantee it. This is evident in MinP: since there is no metaphysical restriction on the form of laws,  if laws are entirely mind-independent primitive facts, how can we determine which propositions correspond to them? However,  a similar problem arises in BSA.  This claim may surprise some philosophers, as BSA is often thought to have an epistemic advantage over non-Humean accounts like MinP,  because it brings laws closer to us by defining them in terms of the mosaic, which is all we can empirically access \citep{earman2005contact}. 

The problem is that \textit{we are not directly given the mosaic}. Just like physical laws, the mosaic postulated in modern physics is a theoretical entity not entailed by direct observations. Our beliefs about its precise nature---such as the global structure of spacetime, its microscopic constituents, and the exact matter distribution---are just as theoretical and inferential as our beliefs about physical laws. They are all parts of a broader theory about the physical world. Just as  MinP require an additional epistemic principle to infer the laws, BSA requires a parallel principle to infer the character of the mosaic. In BSA, this principle becomes a strong epistemic guide concerning  what we should expect about the best system given \textit{our limited evidence}, which itself neither fully determines the mosaic nor the best system. 

After all, in BSA, laws are not summaries of our observations only but of the entire spacetime mosaic---the totality of microphysical facts, only a small fraction of which appear in our macroscopic observations. The optimal true summary depends on the entire mosaic, a theoretical entity. (For this reason, BSA should not be mistaken for strict empiricism.)  In contemporary physics, our best guide to the mosaic is our best guess about the physical laws.  \x{At the end of the day, both MinP and BSA require a super-empirical epistemic principle to infer physical laws.}  In neither case does this principle follow from the metaphysical posits about what laws are. This has implications for the debate between Humeans and non-Humeans, which will be discussed in \S5.3. 

To sharpen the discussion, let us suppose, granting Lewis's assumption of the kindness of nature \cite[p.479]{LewisHSD},  that given a mosaic $\xi$ there is a unique best system whose axioms express the fundamental law $L$: 
\begin{equation}
	L = BS(\xi)
\end{equation}
where $BS(\cdot)$ is the function mapping a mosaic to its best-system law.\footnote{Pragmatic Humeanism might suggest using an alternative best-system function, $BS'(\cdot)$, that selects the system that is ``best for us.'' } Let us stipulate that for both BSA and MinP,  physical reality is described by a pair $(L, \xi)$. In both frameworks, we require that $\xi \in \Omega^L$, where $\Omega^L$ is the set of mosaics compatible with $L$,  meaning that $L$ is true at $\xi$.  In BSA, we also have $L = BS(\xi)$. Thus, in a sense, all we need in BSA is $\xi$; $L$ is not ontologically extra. However,  it does not imply that BSA and MinP are relevantly different in their treatment of epistemic realism. 

%If our evidence consists entirely in observed features of $\xi$, and assuming  not only $L = BS(\xi)$ in BSA but also we humans can compute $BS(\xi)$, then it seems that we can be certain about $L$ in BSA. In contrast, we cannot be certain about $L$ in MinP, since it is possible that $L \neq BS(\xi)$. But the alleged epistemic advantage is an illusion, as the above reasoning does not point to any actual epistemic advantage of BSA over MinP. This is because we are not given $\xi$. The nature of $\xi$ is part of a theory about the physical reality $(L, \xi)$. Hence, we can also be uncertain about $L$ in BSA. 

Let $E$ represent our total empirical evidence---our actual observational data about physical reality. To be generous, let us include not just current data but all past and future data about the universe that we actually gather.  Two key features of $E$ are: 
\begin{itemize}
  \item $E$ does not uniquely determine $\xi$. Multiple candidate mosaics $\xi$ are compatible with the same $E$. (After all, $E$ is a spatiotemporally partial and macroscopically coarse-grained description of $\xi$.) 
  \item  $E$ does not uniquely determine $L$.  Multiple candidate laws $L$ are compatible with the same $E$. (In BSA, this follows from the first point; in MinP, this is even clearer since $L$ can vary independently of $\xi$, up to a point.)
\end{itemize}
Thus, in BSA, just as in MinP, $E$ does not uniquely determine $(L, \xi)$. There is a gap between our empirical evidence and the laws. Ultimately, the gap can be bridged by appealing to simplicity and other super-empirical virtues as epistemic guides. Recognizing the size of this gap highlights the substantial role simplicity and similar principles must play in our reasoning.\footnote{It is useful to contrast this setup with the influential framework suggested by \cite{hall2009humean, hall2015humean}.  Hall introduces the idea of a Limited Oracular Perfect Physicist (LOPP) who has access to the full mosaic $\xi$ as her evidence and nothing else. Her evidence $E_{\text{LOPP}}$ contains vastly more information than $E$, our actual total evidence.  In BSA, if we assume $E_{\text{LOPP}}$ is exhaustive of the entire mosaic (via a ``that’s all'' clause), it uniquely determines $(L, \xi)$.  However, $E_{\text{LOPP}}$ is just as theoretical for Humeans as it is for non-Humeans. A Humean’s best guess about $E_{\text{LOPP}}$ depends on their expectations about what $L$ looks like given $E$.  This further underscores that both BSA and MinP require epistemic principles beyond empirical data to infer laws. } 

\subsection{Empirical Equivalence}

The epistemic gap can be further illustrated through cases of empirical equivalence.  If different laws yield the same empirical evidence, it becomes puzzling how we can be epistemically justified in choosing one over its empirically equivalent rivals---unless we invoke substantive assumptions beyond the metaphysical posits of nomic realism. Discussions in the literature (e.g., \cite{kukla1998studies})  have proposed methods to algorithmically generate empirically equivalent rival laws.  However, some of these resemble Cartesian skeptical scenarios \citep{sep-scientific-underdetermination}, such as the evil-demon hypothesis. In contrast, I offer three new algorithms, modeled on concrete proposals considered in recent philosophy of physics. These algorithms have more limited scopes and should be less controversial. 

 I adopt a fairly weak notion of empirical equivalence:  $L_1$ and $L_2$ are empirically equivalent with respect to actual evidence $E$ if $E$ is compatible with both $L_1$ and $L_2$. This criterion, which emphasizes actual data $E$, is weaker than the notion requiring two laws to agree on all possible data---including data that can in principle be measured in any nomologically possible world. I use this criterion for two reasons. First, it suffices to illustrate the epistemic gap. Second, what is in-principle measurable in the actual world and in nomologically possible worlds depends on what the laws are. Using actual data ensures a more neutral comparison of different hypotheses about laws.\footnote{For those considering the stronger criterion of empirical equivalence, the arguments below can be adapted accordingly. If we consider all possible data, we should also introduce probability distributions over models within the theory and compare the likelihoods of $E$ given different theories. However, this does not alter the dialectic, as there exist probability distributions that assign equal likelihood to $E$ across empirically equivalent laws.} 
 \\

\noindent
\textbf{Algorithm A: \x{``Moving''} parts of ontology (what there is in the mosaic) into the nomology (the package of laws)}. 

   \textbf{General strategy.} This strategy applies to both BSA and MinP. Given a theory of physical reality $T_1=(L, \xi)$, where $\xi$ can be decomposed into two parts $\xi_1$ and $\xi_2$,  we can construct an empirically equivalent rival $T_2 = (L\&\xi_1, \xi_2)$. Here, $\xi_1$ is ``moved'' from  the ontology into the nomology.   One consequence is that empirical evidence from any region of spacetime underdetermines the laws, as it does not determine what belongs in  the ontology versus nomology.%In fact, this algorithm is inspired by recent discussions on quantum Humeanism, where proponents suggest we can move the wave function from the Humean mosaic into the best system as representing a law of nature. However, the basic idea applies to both BSA and MinP. 
 
\textbf{Example: Maxwellian electrodynamics.}  Consider the standard theory of Maxwellian electrodynamics $T_{M1}$: 
  \begin{itemize}
  \item   Nomology: Maxwell's equations, Lorentz force law, and Newton's law of motion.
  \item Ontology: Minkowski spacetime with charged particles $Q(t)$ and an electromagnetic field $F(x,t)$. 
\end{itemize}

Now consider an empirically equivalent rival $T_{M2}$: 
  \begin{itemize}
  \item   Nomology: Maxwell's equations, Lorentz force law, Newton's law of motion, and a highly complex law specifying the exact functional form of $F(x,t)$ in the dynamical equations.
  \item Ontology: Minkowski spacetime with charged particles $Q(t)$, but without an independent electromagnetic field.
\end{itemize}
Our evidence $E$ is compatible with both $T_{M1}$ and $T_{M2}$, as observations will be indistinguishable between them, provided that macroscopic observations register the same particle configurations $Q(t)$ \citep{bell2004speakable}. The additional law in $T_{M2}$ is analogous to the Hamiltonian function in classical mechanics, which encodes force laws. However,  specifying $F(x,t)$ is far more complicated than specifying a Hamiltonian. Both $F(x,t)$ and the Hamiltonian are components of respective laws of nature that tell particles how to move.\footnote{Note that we can  decompose the standard ontology in many other dimensions, corresponding to more ways to generate empirically equivalent laws for a Maxwellian world. This move is discussed at length by \cite{AlbertLPT}. Similar strategies have been considered in the ``quantum Humeanism'' literature. See \cite{miller2013quantum}, \cite{esfeld2014quantum}, \cite{callender2015one, bhogal2015humean}, and \cite{chen2018HU}.  }  \x{Algorithm A illustrates the possibility that we can be mistaken about what the ontology is and what the laws are. Ontology and nomology can be rearranged without altering empirical observations. If $T_{M2}$ is the correct theory, what we commonly believe to be a bit of ontology turns out to be a feature of the laws.}\footnote{\x{I thank an anonymous reviewer for suggesting this clarification.}}   \\

\noindent
\textbf{Algorithm B: Changing the nomology directly}.  
 % in MinP, $L$ constrains $\xi$. in BSA, $\xi$ constrains $L$. 
  
 \textbf{General strategy.}  This strategy applies primarily to MinP.  We generate empirical equivalence by directly modifying  the nomology. Suppose the actual mosaic $\xi$ is governed by the  law $L_1$. Consider $L_2$, where $\Omega^{L_1} \neq \Omega^{L_2}$ and $\xi \in \Omega^{L_2}$.  $L_1$ and $L_2$ are distinct laws because they have distinct sets of \x{mosaics}. Since $E$ (which can be regarded as a coarse-grained and partial description of $\xi$) can arise from both $L_1$ and $L_2$, the two laws are empirically equivalent.  There are infinitely many such $L_2$-candidates, generated by: replacing one non-actual mosaic in $\Omega^{L_1}$ with another outside $\Omega^{L_1}$,  adding mosaics to $\Omega^{L_1}$, or removing non-actual mosaics from $\Omega^{L_1}$.  $L_2$ is empirically equivalent to $L_1$ since $E$ is compatible with both.\footnote{See \cite{manchak2009can, manchak2020global} for more examples.}
 
 \textbf{Example: General Relativity.} Let $L_1$ be the Einstein equation of general relativity, with  $\Omega^{L_1}=\Omega^{GR}$, the set of general relativistic spacetimes. Suppose the actual spacetime is governed by $L_1$, so that $\xi\in \Omega^{L_1}$. Now, consider $L_2$, a law that permits only the actual spacetime and fully specifies its microscopic details, so $\Omega^{L_2}=\{\xi\}.$ Since our evidence $E$ arises from $\xi$, it is compatible with both $L_1$ and $L_2$. However, specifying $L_2$ requires encoding all of $\xi$'s microscopic details, making it (in general) far more complex than $L_1$. ($L_2$ is a case of strong determinism. See \cite{adlam2021determinism} and \cite{chen2022strong, ChenNature2023} for further discussions on strong determinism.)  \\
  
  \noindent
\textbf{Algorithm C: Changing the nomology by changing the ontology}.  
  
 \textbf{General strategy.}  This strategy is tailored to BSA. Since laws in BSA are determined by the mosaic, we can change the nomology by altering the ontology (mosaic). Suppose the actual mosaic $\xi$ is optimally described by the actual best system $L_1=BS(\xi)$. We can construct a different mosaic   $\xi'$, such that it differs from $\xi$ in some unobserved spatiotemporal region,  yet $E$ is compatible with both. Alternatively, we can expand $\xi$ to  $\xi' \neq \xi$ such that $\xi$ is a proper part of $\xi'$. There are infinitely many such $\xi'$-candidates whose best system $L_2=BS(\xi')$ differs from $L_1$. 

 \textbf{Example: General Relativity.} Let $L_1$ be the Einstein equation, and suppose $\xi$ is globally hyperbolic and optimally described by $L_1$, so that $L_1=BS(\xi)$. Now, consider an alternative $\xi'$, which differs only in the number of particles in an unobserved region of a distant galaxy (not in $E$). Since the number of particles is an invariant property of general relativity, it is left unchanged after a ``hole transformation'' \citep{sep-spacetime-holearg}.  Since $\xi$ is globally hyperbolic, we can use determinism to deduce that $\xi'$ is incompatible with general relativity, so that $L_1 \neq BS(\xi')$. Let $L_2$ denote $BS(\xi')$. $L_1\neq L_2$ and yet they are compatible with the same evidence $E$.  Since $\xi'$ violates the conservation of number of particles, $L_2$ should be more complicated than $L_1$. \\

We have considered three algorithms that establish the existence of empirically equivalent rival laws for a world like ours. Moreover, we can combine these methods to construct even more sophisticated equivalences.\footnote{For example, in certain settings, we can change both the ontology and the nomology  to achieve empirical equivalence. For every wave-function realist theory, there is an empirically equivalent density-matrix realist theory \citep{chen2019quantum1}. Their ontology and nomology are different, yet no experiment can distinguish them. } These algorithms draw from recent discussions in the philosophy of physics and do not rely on Cartesian skepticism. Yet, in both BSA and MinP, our evidence underdetermines the laws. If such algorithms are viable, how can we uphold epistemic realism? This brings us to a crucial question for nomic realism: 
\begin{description}
  \item[Puzzle about Nomic Realism:] In such cases of empirical equivalence, what justifies accepting one candidate law over another? 
\end{description}

\section{The Principle of Nomic Simplicity}

It has been recognized---correctly, in my view---that nomic realists must invoke theoretical virtues to choose among empirically equivalent laws underdetermined by evidence. \x{One important example is the \textit{principle of simplicity} (PS), according to which simplicity is a guide to truth and can serve as a tie-breaker among empirically equivalent laws. However,  PS faces a problem of coherence (\S3.1). I propose a better alternative---the \textit{principle of nomic simplicity} (PNS) (\S3.2)---that better aligns with nomic realism. I then explain  how PNS resolves empirical underdetermination  (\S3.3) and generalize its core idea in three ways (\S3.4).}  

\subsection{The Problem of Coherence}

The \textit{principle of simplicity} (PS) has strong intuitive appeal. Paradigm examples of physical laws are strikingly simple, often simpler than alternative laws that yield the same data. Moreover, in the cases of empirical equivalence discussed in \S2.3, the simpler law often seems like the better candidate.

What does it mean for simplicity to be a guide? A guide is not a guarantee. Inferences made under  uncertainty---even when epistemically justified---are fallible.  We might mistakenly regard a simpler law as true when, in fact, the actual physical laws are more complex. A realist should acknowledge this possibility. Indeed, a hallmark of realism is that we can be wrong, even when following rigorous scientific methodology.

This uncertainty can be formulated \x{with epistemic probabilities}:  
\begin{description}
  \item[Principle of Simplicity (PS)] Other things being equal, simpler propositions are more likely to be true. More precisely, other things being equal, for two propositions $L_1$ and $L_2$, if $L_1>_S L_2$, then $L_1 >_P L_2$, where $>_S$ denotes the comparative simplicity, $>_P$ denotes  \x{comparative epistemic prior probability}.\footnote{It may be too demanding to require a total order that induces a \textit{normalizable} probability distribution over the space of all possible laws.  This version in terms of comparative probability is less demanding.}
\end{description}

PS regards simplicity as a guide to \textit{truth}: a simpler proposition is more likely to be true than a more complex one. This interpretation aligns with the usual epistemic gloss on simplicity. However, it is ultimately unsuitable for nomic realism. 

 PS faces an immediate challenge---the problem of nested theories, also known as the problem of conjunctive explanations.\footnote{The problem is widely discussed in philosophy of science but less so in foundations of physics. It was first raised by \cite{popper2005logic} against the Bayesian approach of \cite{wrinch1921xlii}. For recent discussions, see \cite{sober2015ockham, schupbach2017hypothesis} and \cite{henderson2022inference}.} Specifically,  PS leads to probabilistic incoherence, which we shall refer to as the problem of coherence.  

\begin{description}
  \item[Problem of Coherence] PS  leads to probabilistic incoherence.
\end{description}

\begin{figure}
\centering
\begin{tikzpicture}
\coordinate (O) at (0,0);
\draw[thick] (O) circle (2.5);
\draw[thick] (O) circle (1.5) node[right] {$L_1$};
  \fill (2.5, 1) circle (0pt) node[right] {$L_2$};
%\draw (O) circle (0.5);
\end{tikzpicture}
\caption{$L_1$ cannot be more likely to be true than $L_2$, since every model of $L_1$ is a model of $L_2$.}
\end{figure}

Whenever two theories have nested sets of mosaics, such that $\Omega^{L_1} \subset \Omega^{L_2}$,  the probability that $L_1$ is true cannot exceed the probability that $L_2$ is true (Figure 1). For a concrete example from spacetime physics, consider: 
\begin{itemize}
  \item Let $\Omega^{GR}$ denote the set of mosaics compatible with the fundamental law in general relativity---the Einstein equation.
  \item Let $\Omega^{GR^+}$ denote a superset of  $\Omega^{GR}$ that includes additional mosaics violating the Einstein equation.
\end{itemize}
Now, assume there is no simple law that generates $\Omega^{GR^+}$. Since the law of $GR$ (the Einstein equation) is presumably simpler than that of $GR^+$, PS would predict that $GR$ is more likely to be true than $GR^+$. However, it violates probability theory: since every model of  $GR$ is also a model of $GR^+$ but not vice versa, the probability of  $GR$ being true cannot exceed the probability of $GR^+$ being true.   This exemplifies the problem of nested theories, as $\Omega^{GR}$ is a proper subset of $\Omega^{GR^+}$, leading to an incoherent probabilistic assignment under PS.

\subsection{The Correct Principle}

 I propose that \textit{simplicity is a fundamental epistemic guide to lawhood}. Roughly speaking, simpler candidates are more likely to be laws, all else being equal.  This principle resolves the problem of coherence and support epistemic realism in cases of empirical equivalence where simplicity is the deciding factor. Specifically, we should accept:
\begin{description}
  \item[Principle of Nomic Simplicity (PNS)] Other things being equal, simpler propositions are more likely to be laws. More precisely, other things being equal, for two propositions $L_1$ and $L_2$, if $L_1>_S L_2$, then $L[L_1] >_P L[L_2]$, where $>_S$ represents the comparative simplicity, $>_P$ represents the \x{comparative epistemic prior probability}, and $L[\cdot]$ denotes the lawhood operator, mapping a proposition to a claim about its status as a law.\footnote{For example, $L[F=ma]$ expresses the proposition that \textit{F=ma is a law}. The proposition \textit{F=ma} itself is what \cite{lange2009laws} calls a ``sub-nomic proposition.'' } 
\end{description}
From the perspective of nomic realism, one can endorse PNS without endorsing PS. Not all facts are laws---laws correspond to a special subset of facts. In BSA, laws are the best-system axioms. In MinP, laws are the primitive facts constraining physical possibilities. 

PNS  resolves the problem of nested theories. Recall the earlier example of $GR$ and $GR^+$. While we may judge the Einstein equation as more likely to be a law, it is less likely to be true than the equations of $GR^+$. The crucial idea is that simplicity does not select for truth in general, but for truth about lawhood---whether a proposition has the property of being a fundamental law. 

Assuming that fundamental lawhood is factive (as granted in both BSA and MinP), we have: lawhood implies truth, i.e.  $L[p] \Rightarrow p$; yet truth does not imply lawhood, i.e.  $p \nRightarrow L[p]$.  This shows that $L[p]$ is logically distinct from $p$, which is the key to resolve the problem of coherence. 

 Under PS,  nested theories are given contradictory probability assignments. If $L_1$ is simpler than $L_2$, PS suggests  $L_1 >_P L_2$. But if $L_1$ and $L_2$ are nested such that $\Omega^{L_1} \subset \Omega^{L_2}$,  probability theory requires  $L_1 \leq_P L_2$. Contradiction! 

\begin{figure}
\centering
\begin{tikzpicture}
\coordinate (O) at (0,0);
\coordinate (A) at (-0.5, 0.5);
\coordinate (B) at (2, 0.3);
\draw[thick] (A) circle (0.7);
\draw[thick] (B) circle (0.15);
  \fill (-1, 0.5) circle (0pt) node[right] {$L[L_1]$};
    \fill (1.7, 0.9) circle (0pt) node[right] {$L[L_2]$};
    \draw[thick][->]  (2.3, 0.7) -- (B);
\draw[thick] (O) circle (1.5) node[right] {$L_1$};
\draw[thick] (O) circle (3);
  \fill (2.9, 1) circle (0pt) node[right] {$L_2$};
%\draw (O) circle (0.5);
\end{tikzpicture}
\caption{$L_1$ and $L_2$ are nested, while $L[L_1]$ and $L[L_2]$ are not. In this model, $L[L_2] <_P L[L_1] \leq_P L_1 \leq_P L_2$.}
\end{figure}

 Under PNS, the contradiction disappears, because \textit{more likely to be a law} does not entail \textit{more likely to be true}.  If $L_1$ is simpler than $L_2$, but  $\Omega^{L_1} \subset \Omega^{L_2}$, then probability theory still requires $L_1 \leq_P L_2$. However, lawhood operates differently:  $L[L_1] >_P L[L_2]$ can hold independently of the truth probabilities (see Figure 2 for a visualization). Thus, we obtain the following coherent probability chain: 
 \begin{equation}
 L[L_2] <_P L[L_1] \leq_P L_1 \leq_P L_2
\end{equation}
This successfully solves the nested-theories problem and preserves coherence.\footnote{My solution in the context of laws is, in some respects, similar to the solution proposed by \cite{henderson2022inference}, which is based on a ``generative view'' of scientific theories. Henderson suggests that theories with nested sets of mosaics encode different schemas or general principles that generate different sets of specific hypotheses, rendering them mutually exclusive. Henderson focuses on causal model selection and curve fitting, but it would be fruitful to explore connections between our approaches.} \x{In \S3.3, we apply PNS to break ties among empirical equivalent laws. In \S3.4, we generalize PNS to other theoretical virtues and apply its solution to the problem of coherence to non-nested theories.}

\subsection{Simplicity as a Tie Breaker}

PNS is useful in resolving cases of empirical equivalence generated by Algorithms A-C in \S2.3. 

With Algorithm A, $T_2$ generally requires far more complicated laws than $T_1$. For example,  in the case of Maxwellian electrodynamics, the laws of $T_{M2}$ must specify $F(x,t)$ in full detail. Since $F(x,t)$ is not a simple function of space and time, the laws of $T_{M2}$ lack simplicity. By contrast, $T_{M1}$ does not require such a detailed specification. PNS suggests that, all else being equal, we should choose $T_{M1}$ over $T_{M2}$. In a Maxwellian world, we should posit fields in the ontology rather in the nomology.\footnote{Earlier versions of quantum Humeanism with a universal wave function resemble $T_{M2}$, violating PNS. However, the version proposed in \cite{chen2018HU} avoids this issue, as its initial density matrix is as simple as the Past Hypothesis. } 

With Algorithm B, $L_2$ is generally be more complex than $L_1$,  especially if $\Omega^{L_2}$ is obtained from $\Omega^{L_1}$ by adding or removing a few mosaics. For example, a strongly deterministic theory of a sufficiently complex general relativistic spacetime must specify its exact details, requiring laws far more complicated than the Einstein equation.  PNS suggests that other things being equal, we should choose the Einstein equation over such strongly deterministic laws.\footnote{However, not all strongly deterministic theories are excessively complex. See \cite{chen2022strong} for a simple candidate theory that satisfies strong determinism. } 

With Algorithm C, even if the mosaics of $L_1$ and $L_2$ are similar, a simple system like $L_1$ will generally have a more complex counterpart in $L_2$. In fact, if the mosaic deviates significantly from the actual one, there may be no optimal system that simplifies it into a good system. 

 PNS should be distinguished from the simplicity criterion in the Humean best-system account of lawhood (\S5.3). These are conceptually different kinds of principles: the Humean criterion is a metaphysical definition of what laws are, whereas PNS is an epistemic principle guiding ampliative inferences based on total evidence. Even if a Humean expects that the best system is no more complex than the mosaic, it does not follow that they should expect that the best system is relatively simple---because there is no metaphysical guarantee that the mosaic itself is ``cooperative.'' To illustrate, consider an analogy. Suppose we are told that Alice is no shorter than Bob. It does not follow that Alice is tall.  To estimate Alice's height, we need more information about Bob's height. 
 
 Moreover, Algorithm A provides examples, such as $T_{M1}$ and $T_{M2}$, where Humeans cannot distinguish them based on local or global evidence alone. Even in determining what the actual evidence pertains to and whether the local mosaic consists of just particles or both particles and fields, Humeans must appeal to a principle like PNS. Both Humeans and non-Humeans face epistemic uncertainty about the laws and require an additional principle to justify epistemic realism.  If Humeans are epistemically warranted in making such a posit,  so are non-Humeans. 

\subsection{Generalizations}

The Principle of Nomic Simplicity (PNS) can be generalized in several ways. First,  simplicity need not be the only fundamental epistemic guide to lawhood. Other theoretical virtues, such as informativeness and naturalness, can play similar roles. A simple equation that describes too little or does so in overly gruesome terms is unlikely to be a law. This motivates a more general principle:  
\begin{description}
  \item[Principle of Nomic Virtues (PNV)] 
  For two propositions $L_1$ and $L_2$, if $L_1>_O L_2$, then $L[L_1] >_P L[L_2]$, where $>_O$ represents the overall comparison based on theoretical virtues and their tradeoffs, with $>_S$ a contributing factor,  $>_P$ represents  comparative prior epistemic probability, and $L[\cdot]$ denotes \textit{is a law}, an operator mapping a proposition to one about lawhood.
\end{description}
Since $>_O$ does not necessarily induce a total order of all  candidate laws, neither does  $>_P$.\footnote{Nevertheless, in the cases of empirical equivalence discussed in \S2.3, clear winners emerge in terms of overall comparison. } Determining what is ``overall better'' is a holistic matter that requires balancing multiple theoretical virtues, such as simplicity, informativeness, and naturalness. PNV should therefore be seen as a more general epistemic principle than PNS. \x{(For an application of PNV, see footnote \#26.)}

\x{Second, in explanatory contexts where physical laws are not postulated, a further generalization is possible:}
\begin{description}
  \item[Principle of Explanatory Virtues (PEV)] 
  For two propositions $L_1$ and $L_2$, if $L_1>_O L_2$, then $Exp[L_1] >_P Exp[L_2]$, where $>_O$ represents the overall comparison based on theoretical virtues and their tradeoffs, with $>_S$ a contributing factor,  $>_P$ represents the comparative prior epistemic probability, and $Exp[\cdot]$ denotes \textit{is an explanation},  an operator  mapping a proposition to one about explanation.
\end{description}
Epistemic guides for lawhood resemble criteria for inference to the best explanation (IBE).  Selecting a law based on nomic virtues is akin to choosing an explanation based on IBE. \x{(See also footnote \#36.)} %The earlier approach to the problem of coherence can be adapted to solve a similar problem on IBE. 

Finally, the problem of nested theories and its solution can be further generalized.\footnote{I thank Ned Hall for suggesting this generalization.} Consider two hypotheses $L_1$ and $L_2$ that \textit{do not have nested sets of mosaics}. Suppose  that   $L_1$ better balances theoretical virtues than $L_2$. If we naively conclude that $L_1$ is more likely to be true than $L_2$, it can still lead to probabilistic incoherence. To see why, consider the disjunction $L_1 \vee Q$, where $Q$ is much worse than $L_2$, making $L_1 \vee Q$ a poorer overall candidate than $L_2$.    Applying the naive principle again, $L_2$ is more likely to be true than $L_1 \vee Q$. Yet, probabilistic coherence demands the opposite, since $L_2 <_P L_1 \leq_P L_1 \vee Q$.  The root of the problem is that ``more likely to be true'' transmits under entailment, but ``more theoretical virtuous'' does not. That $L_1$ is theoretically better than $L_2$ does not imply that  $L_1 \vee Q$ is theoretically better than $L_2$. In contrast, PEV avoids this mismatch, since  ``more likely to be an explanation'' does not transmit under entailment. Even if $L_1$ is more likely to be an explanation (for the target phenomenon) than $L_2$,  $L_1 \vee Q$ is not necessarily more likely to be an explanation than $L_2$. Thus, it remains probabilistically coherent to endorse:
 \begin{equation}        
Exp[L_1 \vee Q] <_P Exp[L_2] <_P Exp[L_1] \leq_P L_1 \leq_P L_1 \vee Q
\end{equation}
As Ned Hall insightfully observes, the strategy is available whenever we encounter epistemically significant features that do not transmit under entailment. We should connect such features not to the likelihood of truth, but to something else---such as lawhood (PNS/PNV) or explanatory power (PEV). While I primarily focus on PNS, the following discussion also applies to PNV and PEV.

There are further questions about PNS, which I will revisit in \S5. In the next section, I discuss five additional theoretical benefits of PNS, providing further support for its adoption.

\section{Theoretical Benefits}

To further illustrate the theoretical benefits of PNS, I discuss five key issues relevant to nomic realists: induction, symmetries, dynamics, determinism, and explanation. Accepting PNS provides a systematic framework for addressing these issues.

\subsection{Induction}

On nomic realism, Hume's problem of induction\footnote{For an updated review, see \cite{sep-induction-problem}.} is closely related to the problem of underdetermination. We seek to determine physical reality $(L, \xi)$. Given our limited evidence about part of $\xi$ and some aspect of $L$, how can we rationally infer other aspects of $\xi$ or  $L$ that will be revealed in future or possible observations? Without prior knowledge of $(L, \xi)$ and with only limited evidence, we seemingly have no epistemic justification for favoring one possibility over another (\S2.3). On a given law $L$ we know what kind of mosaic $\xi$ to expect. But we are given neither $L$ nor $\xi$.  Without further assumptions, it seems difficult to make sense of the epistemic justification of induction.

Hume connects induction to a principle of uniformity: 
\begin{quotation}
  if Reason determin’d us, it would proceed upon that principle \textit{that instances, of which we have had no experience, must resemble those, of which we have had experience, and that the course of nature continues always uniformly the same}. (\textit{A Treatise of Human Nature}. 1.3.6.4)
\end{quotation}
We label the principle as: 
\begin{description}
  \item[Principle of Uniformity (PU)] Nature is uniform. 
\end{description}
Hume sometimes paraphrases PU as the expectation that ``[from] causes which appear similar we expect similar effects'' or that ``the future will be conformable to the past'' (\textit{An Enquiry Concerning  Human Understanding}, 4.2). 
One interpretation of his argument is that since induction presupposes PU, and PU lacks non-circular justification, induction itself lacks non-circular justification.\footnote{Unlike \cite{ArmstrongWIALON} or \cite{foster2004divine}, I do not suggest that some versions of nomic realism escape Hume's argument.} A potential response is to postulate a fundamental principle like PU that justifies induction but requires no further justification. 

However, does PU provide the right epistemic foundation for nomic realism? Examining three possible interpretations of PU, we find that it is unsuitable for induction, and PNS is a superior alternative.
 
 (1) \textbf{Uniformity of Evidence}. Suppose PU demands that evidence $E$ be uniform, i.e., given the same experimental setup,  outcomes must always be the same. That is not always useful for induction. Experimental setups are never exactly identical, and neither are their outcomes. They are similar in some respects but not others.  Moreover, our evidence for physical theories is highly diverse, arising from different experiments and observations that cross-check the same theory. For example,  evidence for general relativity comes from various sources, including gravitational lensing, time dilation, and gravitational waves \citep{misner1973gravitation, thorne2017modern}.  It is beneficial that they are not all of the same type. 
 %this needs to be clarified or reworded. 

 (2) \textbf{Uniformity of the Mosaic}. Alternatively,  PU may require the mosaic $\xi$ to be uniform. However, this is demonstrably false. The universe we inhabit is highly non-uniform---it contains diverse objects, properties, and structures. Our spacetime region differs drastically from regions with colliding stars or black hole mergers. The Earth’s surface is radically different from even a nearby patch---the Sun’s core, where nuclear fusion converts hydrogen into helium. Despite this non-uniformity, induction remains rational.  In fact, thermodynamic non-uniformity is arguably necessary for the observed temporal asymmetries in our universe, which may be a precondition for induction.\footnote{See \cite{albert2000time}. See also \cite{wallace2010gravity} and \cite{rovelli2019past} for the importance of the hydrogen-helium imbalance in the early universe to the existence of the relevant thermodynamic asymmetries.}

  Moreover, Algorithm A in \S2.3 shows that neither version of PU is sufficient for inductive learning about physical reality. Even if observations and physical phenomena were perfectly uniform throughout the universe, we still would not know what to infer from actual evidence. Observations of the mosaic do not directly reveal its true nature, as illustrated by cases like $T_{M1}$ and $T_{M2}$. For example, even if we observe pointer readings suggesting values for ``electromagnetic fields'' in a given region, this does not automatically warrant the conclusion that electromagnetic fields actually exist there. If we do not know what is revealed by actual observations, we likewise cannot reliably extrapolate from possible observations in unobserved regions.

 (3) \textbf{Uniformity of Laws}. Finally, suppose PU demands the uniformity of law $L$. This interpretation shifts the focus from the mosaic to the law. However, this is also problematic. Some take uniformity of $L$ to mean that laws must take the form  ``for all $x$, if $Fx$ then $Gx$,'' which is a regularity, i.e. a universally quantified statement about the mosaic, holding for everything, everywhere, and everywhen. However, any statement can be rewritten as a universal quantification, making this interpretation vacuous. That I have five coins in my pocket on January 1, 2024 is equivalent to the statement that, for everything and everywhere and everywhen, I have five coins in my pocket on January 1, 2024. Suppose instead $L$ cannot reference particular individuals, locations, or times. This is no longer vacuous, but is too restrictive. Many good candidate laws do refer to particular facts, such as the Past Hypothesis in statistical mechanics, the quantum equilibrium distribution in Bohmian mechanics, the Weyl curvature hypothesis in general relativity, and the No-Boundary Wave Function in quantum cosmology (\S4.3). These laws can be accepted on scientific and inductive grounds, and may be required to ultimately vindicate our inductive practice. Alternatively, suppose PU requires that the same law applies everywhere in spacetime. This is again vacuous, as even an intuitively non-uniform law can be rewritten as a uniform law with temporal variation, such as:  
 \begin{equation}\label{E1}
  F=ma \text{ for } (-\infty, t] \text{ and } \x{F=\frac{1}{7}m^5a \text{ for } (t, \infty)}
\end{equation}
where $F$ is given by Newtonian gravitation and $t$ is a time in the far future. The disjunctive law applies everywhere in some spacetime but is clearly non-uniform in content. Thus, uniformity of laws is not the right foundation for inductive reasoning.
 
Under various interpretations, PU fails to provide a satisfactory foundation for inductive learning about $(L, \xi)$. In contrast, PNS offers a more principled approach for nomic realists. PNS allows us to rationally prefer $F=ma$ over the disjunctive law in (\ref{E1}) when evidence underdetermines them.  What induction ultimately requires is the reasonable simplicity of physical laws, and a simple law may well give rise to a complicated mosaic with an intricate matter distribution. PNS accommodates laws about boundary conditions and particular individuals.  Some simple laws may even have spatiotemporal variations, such as a time-dependent law of motion $F=\frac{1}{t}ma$.  As long as such variations do not require excessive complexity, we can still inductively learn about physical reality based on available evidence, even in a non-uniform spacetime with dramatically different events in different regions.  

PNS thus succeeds where PU fails---it allows us to rationally and inductively learn about  $(L, \xi)$ without  imposing arbitrary or vacuous constraints.  If laws are simple, they may be completely uniform in space and time or else \x{have a simple spatiotemporal dependence (see equation (\ref{N4}) in \S4.2 for a realistic case)}. Such simple laws are empirically discoverable even with finite and limited evidence. PU, understood as a requirement for uniform laws, can be seen as a special case of PNS. Thus, I propose that PNS---rather than PU---should serve as a fundamental epistemic principle underlying inductive learning about physical reality. On nomic realism, justifying induction partially reduces to justifying our acceptance of simple laws.\footnote{While PNS provides a foundation for induction, it is not sufficient on its own. Other theoretical virtues, such as informativeness and naturalness, also play crucial roles. For example, in light of Goodman's new riddle of induction \citep{GoodmanFFF}, we may prioritize simple hypotheses that are formulated in natural terms. Thus, nomic realists should regard  \x{The Principle of Nomic Virtues (PNV)} as the more comprehensive foundation  for induction, with simplicity playing a key, but not exclusive, role. \x{Moreover, we may need to assume that our spatiotemporal location is not exceptionally special; see \cite{schwarz2014proving}.} } 
%(We return to this point in \S5.1.) 

 %One way of thinking about Hume's problem of induction is that we need to adopt an epistemic, non-structural principle. PNS does the job better than the oft-cited PU. What ultimately backs induction is (our rational belief in) the simplicity of physical laws.  Mention PNV? 

\subsection{Symmetries}

Symmetry principles play an important role in theory construction and discovery. Physicists often use symmetries to justify or guide their physical postulates. However, whether symmetries hold is an empirical matter, not guaranteed \textit{a priori}.  Why, then, should we regard symmetry principles as useful, and what do they target? I propose that certain applications of symmetry principles serve as defeasible guides for identifying simple laws. In such cases, their epistemic value is parasitic on that of simplicity.\footnote{For a related perspective, see \cite{north2021physics}.}

%We know that the universe we live in is not completely symmetric or invariant under all transformations. Rather, there is a small set of transformations we regard as physically important: time translation, spatial translation,  and so on. And even these transformations are not strictly obeyed on a global scale, such as in GR.  Their existence is not required for the universe to be friendly to induction and not required for a good scientific theory.

Consider again the toy example in (\ref{E1}), which violates time-translation and time-reversal invariance. By contrast, a better law that upholds both symmetries is: 
 \begin{equation}\label{E2}
  F=ma \text{ for all times}  
\end{equation}
The presence of these two symmetries in (\ref{E2}) and their absence in (\ref{E1}) suggest that, all else being equal, we should prefer  (\ref{E2}) to  (\ref{E1}). This preference can be explained by their relative complexity: (\ref{E2}) is much simpler than (\ref{E1}), and the presence of symmetries serves as an indicator of simplicity.  However, this preference assumes that both equations are compatible with the available evidence (evidence obtained so far or total evidence that will ever be obtained). This does not preclude that \textit{if} empirical data is better captured by (\ref{E1}), then we should prefer (\ref{E1}) to (\ref{E2}).

When symmetry principles function as guides to simplicity, they do so in a defeasible manner. Symmetry principles are not ultimate criteria for theory choice. To illustrate this point, I present two further examples demonstrating that widely accepted symmetry principles are not sacrosanct but rather contingent indicators of simplicity that can be disregarded if a superior, simple theory is available.

%Examples include:
%\begin{itemize}
%\item the invariance-based argument for laws about spacetime structure. 
%\item the invariance-based argument for the Statistical Postulate in Boltzmannian statistical mechanics. 
%  \item the symmetry-guided construction of the  guidance equation in Bohmian mechanics.
%\end{itemize}

%However, they are only defeasible guides to simplicity, and they are not an end in itself. To see this, notice that we may have perfectly attractive theories that do not have any familiar symmetries. That in itself is not a regrettable feature, but may be compatible with the ultimate theory we want, namely simplicity and explanatory power. I shall use two examples to illustrate this. 

\begin{figure}
\centerline{\includegraphics[scale=0.4]{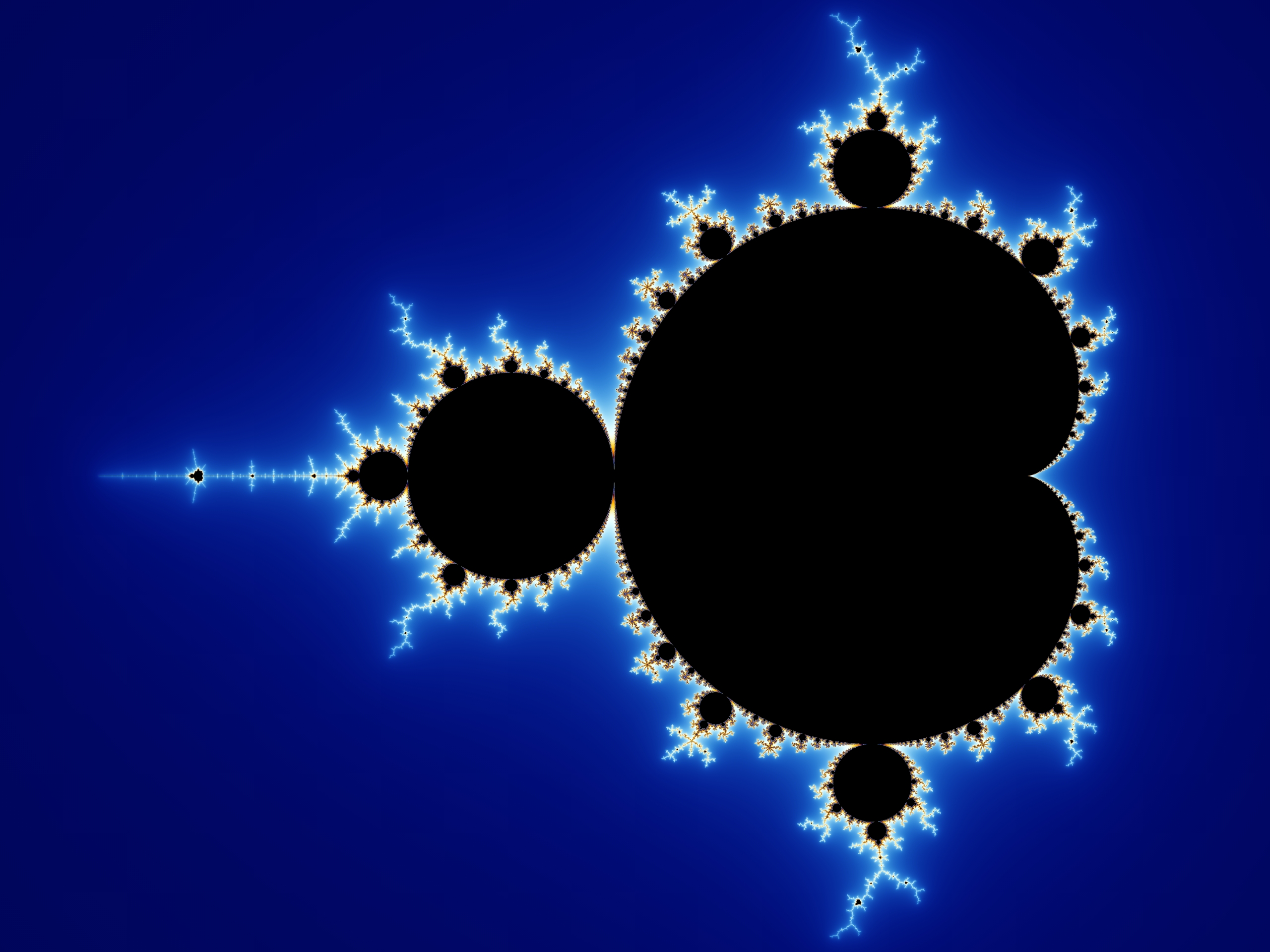}}
\caption{The Mandelbrot set with continuously colored environment. Picture created by Wolfgang Beyer with the program Ultra Fractal 3, CC BY-SA 3.0, https://creativecommons.org/licenses/by-sa/3.0, via Wikimedia Commons}
\end{figure}

The first is a toy example of the Mandelbrot world (Figure 3). The Mandelbrot set in the complex plane is generated by the simple rule that a complex number $c$ belongs to the set if the function  
\begin{equation}\label{Mandel}
	f_c (z) = z^2 +c 
\end{equation}
does not diverge when iterated from $z=0$. (For example, $c=-1$ belongs to this set because the sequence $(0,-1,0,-1,0,-1, ...)$ is bounded, whereas $c=1$ does not because the sequence $(0,1,2,5,26,677,458330, ...)$ diverges.  For a detailed description and visualization, see \cite[ch.3-4]{roger1989emperor}.)  The Mandelbrot set exhibits a striking fractal structure, where zooming in reveals self-similar substructures at every level of magnification, yet always with novel elements.

Now, let us endow the Mandelbrot set with physical significance and interpret it as the distribution of matter in a two-dimensional spacetime, which we call the \textit{Mandelbrot world}, $\xi_M$. We stipulate that the fundamental law of the Mandelbrot world is the rule just described, which we denote by $L_M$. The fundamental law is compatible with exactly one world.\footnote{We note that the patterns of the Mandelbrot world are not exceptionally fine-tuned, as they are stable under certain changes to the law. For example, as \cite[p.94]{roger1989emperor} points out, other iterated mappings such as $	f_c (z) = z^3 + iz^2+ c $ can produce similar patterns.  }

The physical reality consisting of $(L_M, \xi_M)$ is conducive to scientific discovery. If we were inhabitants in the Mandelbrot world, we could infer the structure of the entire $\xi_M$ from the structure of its parts by discerning the law $L_M$. However, $L_M$ does not exhibit recognizable spatial or temporal symmetries.\footnote{There is, however, the reflection symmetry about the real axis. But it does not play any useful role here, and we can just focus on the upper half of the Mandelbrot world if needed.} In fact, conventional notions of symmetry do not apply to $L_M$, because it is not formulated as a differential equation. Despite lacking symmetries,  $(L_M, \xi_M)$ is a perfect example of an ultimate theory (though not of the actual world). It is an elegant and powerful explanation for the intricate patterns in the Mandelbrot world. No alternative explanation, even one with more symmetries, would be preferable. Here, symmetry principles are unnecessary for selecting the correct law, as we already have a simple and compelling candidate. The absence of symmetries is not a regrettable feature of the world, but a consequence of its simple law.  

A second, more realistic example is the Bohmian Wentaculus \citep{chen2018IPH, chen2018HU, chen2023detlef}. If we adopt the nomic interpretation of the quantum state, justified by the simplicity of the initial density matrix, then we can understand the mosaic $\xi_B$ as consisting solely of particle trajectories in spacetime.  The fundamental dynamical law $L_B$  is given by: 
 \begin{equation}\label{N4}
\frac{dQ_i}{dt} = 
\frac{\hbar}{m_i} \text{Im} \frac{\nabla_{q_{i}}   W_{IPH} (q, q', t)}{W_{IPH} (q, q', t)} (Q) = 
\frac{\hbar}{m_i} \text{Im} \frac{\nabla_{q_{i}}   \bra{q} e^{-i \hat{H} t/\hbar} \hat{W}_{IPH} ( t_0) e^{i \hat{H} t/\hbar} \ket{q'} }{ \bra{q} e^{-i \hat{H} t/\hbar} \hat{W}_{IPH} (t_0) e^{i \hat{H} t/\hbar} \ket{q'}} (q=q'=Q),
\end{equation}
where the right-hand side represents the canonical formulation of the law. This equation explicitly violates time-translation invariance, since the expression 
$$\text{Im} \frac{\nabla_{q_{i}}   \bra{q} e^{-i \hat{H} t/\hbar} \hat{W}_{IPH} ( t_0) e^{i \hat{H} t/\hbar} \ket{q'} }{ \bra{q} e^{-i \hat{H} t/\hbar} \hat{W}_{IPH} (t_0) e^{i \hat{H} t/\hbar} \ket{q'}}$$ 
  takes on different forms at different times. Despite this, the physical reality described by the Bohmian Wentaculus may be our world, and the equation can be discovered scientifically. The law is a version of the Bohmian guidance equation that directly incorporates a version of the Past Hypothesis. Thus, $(L_B, \xi_B)$ describes a physical reality amenable to scientific discovery yet does not uphold time-translation invariance. 

In the Bohmian Wentaculus world, symmetry principles remain applicable, but the fundamental dynamical law explicitly breaks time-translation invariance. However, this lack of symmetry is not problematic, as we have already identified a simple and viable candidate law. Once again, the absence of symmetry is a consequence of the law's simplicity. In the hierarchy of epistemic criteria, PNS takes precedence over symmetry principles and serves as the deeper justification for theory selection.

\subsection{Dynamics}

We have strong reasons to accept fundamental laws of boundary conditions.  However, not all boundary conditions are suitable candidates for fundamental lawhood. Epistemic guides, such as simplicity, allow us to be selective in postulating boundary condition laws and to sometimes give greater weight to those that incorporate dynamical laws.

To begin, consider some motivations for positing fundamental laws of boundary conditions. Cosmologists have proposed that fundamental physical laws should include a law specifying the universe’s initial condition. This idea is implicit in Hartle and Hawking’s work on the No-Boundary Wave Function, where they postulate a universe (described by a universal wave function) that smoothly contracts to a single point in the past \citep{hartle1983wave}. Moreover, \cite{hartle1996scientific} suggests that the most general laws of physics consist of two elements: a dynamical law governing fundamental interactions and a law specifying the initial boundary condition of the universe—both of which are essential for cosmology.

%summarizes his view as follows:
%\begin{quotation}
% ``A present view, therefore, is that the most general laws of physics involve two elements: 
%  \begin{itemize}
%  \item The law of dynamics prescribing the evolution of matter and fields and consisting of a unified theory of the strong, electromagnetic, weak and gravitational forces.
%  \item A law specifying the initial boundary condition of the universe.
%\end{itemize}
%There are no predictions of any kind which do not depend on these two laws, even if only very weakly, or even when expressed through phenomenological approximations to these laws (like classical physics) appropriate in particular and limited circumstances with forms that may be only distantly related to those of the basic theory.''
%\end{quotation}

A second motivation comes from the indispensable role of probabilistic boundary conditions in the predictive success of certain physical theories \citep{ismael2009probability}. If these conditions underpin objective probabilities in physics, they can rightfully earn the status of fundamental laws. Examples include the Past Hypothesis and the Statistical Postulate of the Mentaculus theory \citep{albert2000time, LoewerCatSLaw}. Another independent case arises in Bohmian mechanics, where the quantum equilibrium distribution asserts that the initial particle configuration is distributed according to $\rho(q,t_0) = |\Psi(q,t_0)|^2$. While this postulate is made plausible by Bohmian dynamics, it is not strictly entailed by it. Nevertheless, it arguably functions as a physical law within Bohmian mechanics \citep{barrett1995distribution, loewer2004david, callender2007emergence}.

%There is an interesting methodological argument for the nomological status of the quantum equilibrium distribution that is implicit in \cite{barrett1995distribution}'s discussion. Barrett shows that the postulate places a very strong constraint on the relationship between the universal wave function and the initial particle configuration and is essential for the empirical adequacy of the theory. At the level of theory construction for a Bohmian theory, there is considerable freedom in our choice of the boundary condition postulate and the dynamical laws. To enforce the Born rule predictions, we can either postulate the probabilistic boundary condition or  tweak the dynamical equations (such as by adding a stochastic element in the guidance equation). Since there is such a freedom in theory construction as we try to capture the empirical data, it seems that the probabilistic boundary condition is playing a similar role as the dynamical equations, namely as physical laws that explain the Born rule statistics.

A common feature of these examples is their simplicity. While many boundary conditions exhibit detailed correlations, only a select few qualify as fundamental laws---namely, those that are both simple and explanatory. One might ask why we prefer the Past Hypothesis, a macroscopic description, over a precise microscopic specification of the universe’s initial condition. The reason is that the Past Hypothesis is far simpler while still providing a powerful explanation for a wide range of temporally asymmetric regularities.

The simplicity of boundary condition laws suggests that dynamical laws are almost inevitable. Scientific explanations of natural phenomena emerge from the interplay between simple boundary conditions and dynamical laws. As a result, dynamical laws carry substantial explanatory weight on their own.

\subsection{Determinism}

Nomic realism is often associated with other reasonable expectations about physical laws. Here, I explore issues related to determinism and superdeterminism. 

  \begin{figure}
\centering	
\begin{tikzpicture}
    \begin{axis}[
     %   axis x line=bottom,
     %   axis y line=left,
        xmin=0, xmax=10, 
        ymin=0, ymax=10,
        ytick=\empty,
        xtick=\empty,
        ]
      %  \draw (axis cs:2,9) to [bend right=30] coordinate[pos=0.2] (l_i) (axis cs:8,5);
       % \fill (l_i) circle (2.2pt) node[above right] {$B$};
     %   \draw (axis cs:2,8) to [bend right=30] coordinate[pos=0.2] (l_i) (axis cs:8,4.5);
                    \draw (axis cs:1,4) to [bend right=20]  coordinate[pos=0] (dl_j) (axis cs:9,0.5); 
                            \fill (dl_j) circle (0pt) node[left] {$X_0$};
        %        \draw (axis cs:1,4.5) to [bend right=20] coordinate[pos=0] (dl_j) (axis cs:9,1);
             %               \fill (dl_j) circle (0pt) node[left] {$X_1$};
            \draw (axis cs:1,5) to [bend right=20] coordinate[pos=0] (dl_j) (axis cs:9,1.5);
                            \fill (dl_j) circle (0pt) node[left] {$X_1$};
     %   \draw (axis cs:.5,5.5) to [bend right=20] coordinate[pos=0.8] (dl_j) (axis cs:9,2);
        \draw (axis cs:1,6) to [bend right=20] coordinate[pos=0] (dl_j) (axis cs:9,2.5);
                                    \fill (dl_j) circle (0pt) node[left] {$X_2$};
    %            \draw (axis cs:.5,6.5) to [bend right=20] coordinate[pos=0.8] (dl_j) (axis cs:9,3);
                \draw (axis cs:1,7) to [bend right=20] coordinate[pos=0] (dl_j) (axis cs:9,3.5);
                            \fill (dl_j) circle (0pt) node[left] {$X_3$};
   %             \draw (axis cs:.5,7.5) to [bend right=20] coordinate[pos=0.8] (dl_j) (axis cs:9,4);
                \draw (axis cs:1,8) to [bend right=20] coordinate[pos=0] (dl_j) (axis cs:9,4.5);
                            \fill (dl_j) circle (0pt) node[left] {$X_4$};
    %            \draw (axis cs:.5,8.5) to [bend right=20] coordinate[pos=0.8] (dl_j) (axis cs:9,5);
                \draw (axis cs:1,9) to [bend right=20] coordinate[pos=0] (dl_j) (axis cs:9,5.5);
                            \fill (dl_j) circle (0pt) node[left] {$X_5$};
%                \draw (axis cs:.5,9.5) to [bend right=20] coordinate[pos=0.8] (dl_j) (axis cs:9,6);
     %   \fill (dl_j) circle (2.2pt) node[above right] {$D$};
    \end{axis}
    \end{tikzpicture}
    \caption{Schematic illustration of a deterministic theory $T$.  $\Omega^T$ contains six nomologically possible worlds that do not cross in state space.}
\end{figure}

Borrowing ideas from  \cite[pp.319-321]{MontagueFP},  \cite[p.360]{LewisNWTU}, and \cite[pp.12-13]{earman1986primer},  I define determinism as follows (also see Figure 4): 

\begin{description}
  \item[Determinism$_{T}$] Theory $T$ is  \textit{deterministic} just in case, for any two \x{mosaics} $w, w'\in \Omega^T$, if $w$ and $w$ agree at any time, they agree at all times. 
\end{description} 
  Intuitively, $T$ is deterministic if and only if no two mosaics compatible with $T$ cross in state space (overlap at any time). 
  
  Using  $\Omega_{\alpha}$ to denote the set of mosaics compatible with the actual law, we can define: 
  \begin{description}
  \item[Determinism$_{\alpha}$]    The actual world $\alpha$ is \textit{deterministic} just in case, for any two \x{mosaics} $w, w'\in \Omega_{\alpha}$, if $w$ and $w'$ agree at any time, they agree at all times. 
\end{description}
 Determinism is true if and only if the actual world $\alpha$ is deterministic in this sense. 

  In MinP, given any mosaic $\xi$, there are many possible choices of $L$ such that $\xi \in \Omega^L$ and nomologically possible mosaics in $\Omega^L$ do not cross. A simple way to construct such an $L$ is as follows: define a two-member set $\Omega^L=\{\alpha, \beta \}$ where $\alpha$ and $\beta$ never agree at any time (or on any Cauchy surface). Any law with this domain satisfies the definition of determinism. As long as $\alpha$ is not a world where every logically possible state of the universe happens some time in the universe,  there are numerous choices of $\beta$ that preserves determinism. Without an additional principle guiding our expectations about  $L$, determinism in MinP becomes almost trivial.\footnote{See \cite{russell1913notion} for a related argument. Algorithm B in \S2.3 provides another example involving strong determinism.}
 
 In contrast, BSA faces the opposite challenge---determinism becomes exceedingly difficult to satisfy. Even if evidence $E$ is optimally summarized by a deterministic law $L$, this does not guarantee (or make likely, without further assumptions) that the entire mosaic is optimally summarized by $L$. Small ``perturbations'' within the mosaic, such as those introduced in Algorithm C in \S2.3, can easily make the best system fail determinism.\footnote{See \cite{builes2022ineffability} for a related argument.}
 
 This raises a key question for nomic realists: what constitutes a principled reason to believe that determinism is neither trivial (in MinP) nor epistemically inaccessible (in BSA)?\footnote{For more general definitions of determinism that extend beyond the traditional structure of ``states at a time,'' see \cite{adlam2021determinism} and \cite{chen2022strong}. The arguments here can be adapted to those frameworks. In BSA, strong determinism is particularly difficult to achieve, as almost any minor perturbation in a strongly deterministic mosaic will render its best system non-strongly-deterministic. Similarly, \textit{delocalized holistic determinism}, as defined by Adlam, is also fragile: small changes to the mosaic can move it outside the class of ``hole-free'' spacetimes. See \cite[\S3.3]{adlam2021determinism} for a relevant construction.}
 
 With PNS, determinism ceases to be trivial in MinP. While any given mosaic $\xi$ may be compatible with multiple deterministic laws, not every mosaic will permit a relatively simple law that is both deterministic and explanatory.  The non-triviality of determinism in MinP corresponds to the non-triviality of finding a law that is simple and deterministic, as that is not guaranteed for every metaphysically possible mosaic. 

Similarly, PNS makes determinism epistemically accessible in BSA. This is connected to induction: we are justified in believing that the best system of the actual mosaic is relatively simple, even though the actual evidence does not entail this. If the actual evidence is best summarized by a deterministic law, then we have epistemic justification for extending this inference beyond our observations, even into regions that will never be observed. That is, we are justified in believing that the entire mosaic, $\xi$, can be summarized by a simple law that happens to be deterministic. 

Related to determinism is the concept of superdeterminism in quantum foundations. A superdeterministic theory is one that is deterministic but violates statistical independence \citep{hossenfelder2020rethinking}. Roughly speaking, a theory violates statistical independence if the probability distribution of fundamental physical variables is not independent of detector settings. Superdeterminism is motivated as a way to evade Bell’s theorem and the implication of non-locality.

Although superdeterministic laws are not metaphysically impossible in either BSA or MinP, PNS provides a principled objection to those laws. The  constraints on empirical frequencies imposed by superdeterminism are extremely severe, making it difficult to express such laws in any simple formulation. Unlike the Past Hypothesis, which can be stated in a relatively simple description of the initial matter distribution or spacetime structure, there is no reason to believe that superdeterministic laws could be simple at all. Given simpler alternatives such as Bohmian mechanics and objective collapse theories, PNS justifies assigning low credence to superdeterministic theories. For a more comprehensive critique of superdeterminism and the role of nomic simplicity, see \cite{ChenBell}.

\subsection{Explanation}

There is a strong connection between nomic realism and scientific explanation. The purpose of postulating laws, whether in BSA or MinP, is to provide scientific explanations. However, not all candidate laws offer the same quality or type of explanation. Consequently, on both versions of nomic realism, we may ask whether there is a principled reason to expect successful scientific explanations for all phenomena. 

In MinP, laws provide good explanations only when they are sufficiently simple.  Constraints, in and of themselves, do not always yield satisfying explanations  \cite[p.45]{chenandgoldstein}.  Many constraints are complicated and thus insufficient for understanding nature.  For example, the constraint given by  $\Omega^L = \{\xi_M\}$, which fully specifies the mosaic, does not sufficiently explain  the pattern in the Mandelbrot world. Knowing why there is a pattern requires more than knowing the exact distribution of matter.  

Although many candidate laws in MinP can constrain the mosaic, not all are simple enough to illuminate its structure. With PNS, we expect the actual constraint to be relatively simple. The Mandelbrot law provides a far superior explanation compared to the constraint $\Omega^L = \{\xi_M\}$, as the former is simple and explanatory while the latter is overly detailed and unilluminating.\footnote{PNS and other epistemic guides may be regarded as responses to questions about primitivism raised by \cite{hildebrand2013can}.}   

In BSA, laws are defined by their role in systematizing the mosaic. However, whether a systematization exists that is simpler than the mosaic itself is a contingent matter, depending on the microscopic and global structure of the mosaic. Not all mosaics support a systematization that unifies diverse phenomena in an illuminating way \citep{LoewerFluke2023}. BSA only guarantees that the best system is no more complex than the full specification of the mosaic. Some mosaics may support no better optimal summary than the exact specification of the mosaic itself. Hence, in BSA, having successful explanations is not automatic. It requires the mosaic to be favorable. 

Certain mosaics in BSA, however, are favorable: they permit optimal summaries that are simpler than themselves and offer ``Humean explanations'' of the mosaic. Yet, most mosaics may not be favorable \citep{lazarovici2020typical}.  There exist mosaics that are underdetermined by actual evidence and lack good summaries. However, with PNS, we are epistemically justified in inferring that the actual best system is relatively simple, enabling it to provide a Humean explanation of the actual mosaic. Essentially, we expect that the actual Humean mosaic is favorable, i.e. it cooperates with our scientific methodology and can be unified under a reasonably simple best system. 

Thus, on both MinP and BSA, the viability of scientific explanation ultimately hinges on PNS. 

\section{Epistemic Fundamentality}

I have argued that  PNS yields substantive theoretical benefits. For that reason, I regard it as a fundamental epistemic principle. In this section, I address three key issues: the problem of justification, the problem of precision, and the epistemology of  laws within both Humean and non-Humean frameworks. 

\subsection{The Problem of Justification}

Unlike logical consistency and probabilistic coherence, PNS is a non-structural epistemic principle. It is neither analytic nor empirically discoverable. Moreover, it does not follow from metaphysical realism that laws must be relatively simple. In both BSA and MinP, laws \textit{can} be extremely complicated.\footnote{If the best system is too unwieldy, Humeans may argue that there is nothing that truly deserves the title of lawhood. } \cite[p.158]{roberts2008law} suggests that a principle like PNS function as a ``synthetic \textit{a priori}'' claim about metaphysically contingent truths---one that is much stronger than what even Kant might have endorsed.  

One might reasonably ask what could justify such a strong principle. It is natural to worry: 
\begin{description}
  \item[Problem of Justification] There is no plausible epistemic justification for the  principle of (nomic) simplicity. 
\end{description}

%I need to connect more directly to the problem of justification as formulated. 
One potential justification for PNS is an argument from reflective equilibrium. There exist numerous cases of empirically equivalent theories where the salient difference between them is their relative complexity. For example, if we are epistemically justified in preferring $T_{M1}$ to $T_{M2}$ because the former has simpler laws, or in accepting $GR$ over an artificially complexified version $GR^+$ for the same reason, the simplicity must function as an epistemic guide to lawhood.\footnote{For a similar argument, see \cite{lycan2002explanation}.} The broader applications of PNS discussed in \S4 further reinforce this idea.
 
 Reflecting on these judgments, we may conclude that simplicity as a guide to lawhood is an epistemic posit we must make in order to sustain epistemic realism about laws. It is what we implicitly assume when we dismiss or assign lower credence to exceedingly complex empirical equivalents. If our preferences in cases of empirical equivalence are to be epistemically justified,  simplicity must function as a legitimate epistemic guide. This suggests that \textit{PNS is not merely a pragmatic principle, even though it carries pragmatic advantages} (e.g., simpler laws are easier to conceive, manipulate, and test). As an epistemic guide, simplicity ultimately aims at identifying certain truths about lawhood and justifying our beliefs in such truths. Rejecting simplicity as an epistemic principle would undermine epistemic realism itself—an option unavailable to nomic realists.

Why should we regard simplicity as a fundamental epistemic guide that requires no further justification? One compelling argument is its role in vindicating induction (\S4.1). Inductive reasoning about physical reality is essential to both scientific practice and nomic realism. We can make a transcendental argument: since science presupposes induction, we must accept its epistemic rationality. However, as Hume famously observed, induction lacks a non-circular justification: any attempt to justify it deductively or probabilistically ultimately relies on premises that themselves require induction. Thus, no justification can fully satisfy the skeptic. At some point, we must adopt fundamental epistemic principles that explain how and why induction works.

I suggest that PNS is a good candidate for such a fundamental posit, though it need not be the only one (see footnote \#26.) If we are rational in believing that physical laws are relatively simple, we can reasonably assume that they are either uniform across space and time or else provide a simple rule that specifies how they change. Using standard scientific methodology, we can then discover physical laws and the natural phenomena they govern.

PNS operates at the right level of generality and makes correct connections to symmetries, determinism, and explanation. While accepting PNS as a fundamental epistemic principle may seem bold, it is worth noting that we already accept similar foundational principles---such as the reliability of perception and the absence of evil demons. PNS is simply another necessary epistemic assumption that allows us to navigate and succeed in our epistemic pursuits.\footnote{\x{  An astute referee notes that my formulation of PNS is \textit{non-veritistic}, meaning  it applies even if physical laws are not actually simple. This contrasts with an alternative approach that postulates a \textit{veritistic} epistemic norm---one that depends on what the world is like. Under that view, if physical laws are simple, then we ought to believe they are probably simple; but if they are complex, then the norm does not hold, and we are not required to believe they are probably simple. A challenge for the veritistic approach is that, unless we independently know whether physical laws are simple, we cannot determine whether to apply the epistemic norm. PNS, by contrast, does not rely on such prior knowledge---it prescribes that simplicity should guide our epistemic commitments regardless. However, this also means that PNS can sometimes misalign with reality: if physical laws are simple, PNS leads us to the right belief; if they are complex, it does not. Since we lack an independent means of verifying the character of physical laws in advance, scientific discovery may not always reflect physical reality with complete accuracy. In this sense, nomic realists who adopt PNS must accept a degree of epistemic fallibility. But that is life---we navigate uncertainty with the best principles available, even if they are not infallible.  }}

 What about attempts that reduce simplicity to structural epistemic principles, such as the likelihood principle? As far as I can see, these reductive approaches do not apply in the cases of empirical equivalence discussed here. For example, the AIC model-selection criterion advocated by \cite{forster1994tell} is designed for predictively inequivalent theories.  \cite{sober1996parsimony} argues that the AIC framework provides a justification for simplicity only when theories make different predictions. He suggests that parsimony considerations have no epistemic force when applied to predictively equivalent theories.
 
 In the absence of a successful reduction of simplicity that resolves cases of empirical equivalence, it remains justified to regard simplicity as a fundamental epistemic guide. Of course, if someone were to provide a proof that simplicity can be reduced to a deeper structural principle, we should remain open to the idea and consider simplicity as derivative. However, the existence of reasonable algorithms capable of generating empirical equivalents casts doubt on the viability of such a reduction. Similar concerns apply to reductive approaches to Inference to the Best Explanation (IBE), such as \cite{henderson2014bayesianism}'s proposal. Henderson argues that explanatory considerations and theoretical virtues may not be necessary for determining epistemic priors, as simpler theories often receive a greater boost from evidence. However, as Algorithms A-C show, we can construct empirical equivalents where the more complex theories assign equal or higher likelihoods to actual evidence, preventing simpler alternatives from receiving a probabilistic advantage. In such cases, assigning higher priors to simpler theories is necessary, reinforcing the need for a fundamental epistemic principle like PNS.

\subsection{The Problem of Precision}

Simplicity is a vague notion. If PNS is to be regarded as a fundamental epistemic principle, this vagueness might seem undesirable. It is natural to worry:

\begin{description}
  \item[Problem of Precision] There is no precise standard of simplicity.
\end{description}

It is unrealistic to demand a single, universal measure of simplicity for physical laws. Simplicity has multiple dimensions, as demonstrated by research in computational complexity, statistical testing, and the philosophy of science. These include factors such as the number of adjustable parameters, the length of axioms, algorithmic compressibility, and conceptual elegance.\footnote{For an overview of these different measures, see \cite{sep-simplicity} and \cite{iep-simplicity}.} Some laws may be more unified in their conceptual framework, excelling in one aspect of simplicity, but require longer formal expressions, making them less simple in another respect. There is no definitive method for weighing these different dimensions against one another. Moreover, not all physical laws are differential equations---some take the form of boundary-condition laws or conservation principles---so expecting a single simplicity criterion to apply universally is unreasonable. A more natural approach is to assess simplicity holistically, considering these multiple aspects together.\footnote{Alternatively, we may treat simplicity as a family of related concepts, with the principle of nomic simplicity understood as a collection of related epistemic principles.}

The vagueness of simplicity might seem problematic for nomic realists, particularly those who rely on PNS. However, what ultimately matters is that there is sufficient agreement on paradigm cases. While hard cases of simplicity comparison exist, there are also clear-cut cases---such as the relationship between $T_{M1}$ and its empirical equivalents generated by Algorithm A, or between general relativity and its empirically equivalent alternatives produced by Algorithms B and C. This situation is reminiscent of Lewis's assumption that Nature is kind to us, such that borderline cases do not arise in realistic comparisons. The vagueness of simplicity is thus no more troubling than the well-known vagueness inherent in the BSA account of lawhood.
   
   Moreover, vagueness does not imply that there are no facts about simplicity comparisons. A useful analogy can be drawn from moral philosophy. Moral judgments are often holistic and vague---there are hard cases where moral considerations conflict, but this does not undermine the existence of clear cases. For example, there is little controversy in judging that helping a neighbor in need is morally better than torturing their cat for fun. Moral realists argue that our robust moral intuitions in paradigm cases are not invalidated by the existence of borderline cases. Similarly, the fact that some cases of simplicity comparison are difficult does not mean there are no objective facts about simpler and more complex theories.

A deeper worry might be that vagueness is a sign of non-fundamentality---that any truly fundamental epistemological principle must be exact. However, there is little reason to accept this assumption. I am not aware of any non-structural epistemic principle that is completely precise. In the case of PNS, we have principled reasons to expect vagueness, and its imprecision is appropriate given its broad range of applications. PNS plays a role in guiding our epistemic commitments regarding induction, symmetries, dynamics, determinism, and explanation. Given this diversity, different measures of simplicity may not always align. Furthermore, if we accept the possibility of fundamental laws that are themselves vague \citep{chen2018NV}, it is natural to expect that the measure of nomic simplicity is also vague.

Another reason to tolerate some vagueness in simplicity comes from its connection with algorithmic randomness, an active area of research in mathematics and computer science. Mathematicians and computer scientists begin with an intuitive, pre-theoretical notion of randomness and develop various formal definitions, some of which prove to be theoretically fruitful. Notable examples include Kolmogorov's incompressibility criterion, Martin-L\"of's effective typicality, and game-theoretic notion of fair gambling \citep{dasgupta2011mathematical}. Remarkably, under idealized conditions, these definitions are provably equivalent, demonstrating that the vague pre-theoretical concept captures a genuine mathematical reality.

However, these formal definitions do not fully eliminate vagueness. When applied to finite mosaics, vagueness re-emerges \cite[p.56]{li2019introduction}. Instead of drawing a sharp boundary between random and non-random sequences, we must adopt a more flexible criterion: a finite sequence is considered random if it cannot be represented by a significantly shorter algorithm. What counts as ``significantly shorter'' remains vague. Yet this does not prevent the legitimate application of randomness to finite sequences. Similarly, we can acknowledge and tolerate some vagueness in simplicity without undermining its epistemic role.

Since algorithmic randomness serves as a measure of complexity, it also provides a useful perspective on simplicity.  A non-random sequence (satisfying certain frequency properties) may be considered simple---for example, the alternating sequence $(010101......)$ is clearly simpler than a random sequence. A non-random mosaic (of a particular type) can be captured by a suitably simple law. This suggests the following duality: 
\begin{description}
  \item[Duality] Simplicity and algorithmic randomness are duals of each other.
\end{description}
Since algorithmic randomness is appropriately vague, simplicity is too. 

\subsection{Humeanism vs. Non-Humeanism}

Does PNS follow from the metaphysical commitments of BSA? The answer is no. Unpacking why this is the case sheds light on a broader debate between Humeans and non-Humeans.
 
 To begin,  recall the comparison between $T_{M1}$ and $T_{M2}$. According to PNS, a Humean scientist living in a world with Maxwellian data should prefer $T_{M1}$ to $T_{M2}$ because the laws of $T_{M1}$ are simpler. However, in BSA, it is metaphysically possible that the actual ontology contains no fields. If that is the case, the best system will correspond to the highly complex laws of $T_{M2}$. This implies that the best system of the mosaic may differ from what we \textit{should} accept as the best system given our evidence. 

 There is no contradiction here because \textit{what the laws are} can differ from \textit{what we should believe the laws are}.  Thus, defenders of BSA find themselves in a similar epistemic position as defenders of MinP. Even if the laws of $T_{M2}$ represents the actual governing laws, a defender of MinP would still, and should still, regard $T_{M1}$ as more likely.  Both Humeans and non-Humeans can be mistaken about physical reality even when acting completely rationally. This is a feature, not a bug---nomic realists must accept that they are fallible.
 
 This observation has implications for a  prominent argument against non-Humeanism. According to an influential view, Humeanism enjoys an epistemic advantage over non-Humeanism because it offers better epistemic access to laws.\footnote{See, for instance, \cite{earman2005contact} and \cite{roberts2008law}. For a related argument concerning the epistemic inaccessibility of dispositional properties, see \cite{SchwarzKnowing2023}.} The argument suggests that because the Humean mosaic is all that we can empirically access---and because laws supervene on the mosaic---Humeanism secures a more direct epistemic link to laws than non-Humeanism, which postulates additional facts about laws that are empirically undecidable.
 
 However, if the analysis in this paper is correct, such arguments are epistemically irrelevant. We never, in fact, occupy a position to observe everything in the mosaic. Our total evidence $E$ neither exhausts the mosaic $\xi$ nor directly reveals the microscopic details of even the region we occupy. If both Humeans and non-Humeans must rely on independent epistemic principles to ensure epistemic access to laws, then Humeanism has no real epistemic advantage. In practice, our access to laws depends on principles like PNS, which do not follow from the metaphysical commitments of either Humeanism or non-Humeanism. In this respect, both views are epistemically \textit{on a par} when it comes to the discovery and evaluation of laws.
  
 The connection between Humeanism and PNS is somewhat indirect. PNS is an epistemic principle that tells us what system we should \textit{believe} given our total evidence, while BSA is a metaphysical account specifying what the best system \textit{is} given the total mosaic. Since BSA asserts that $L = BS(\xi)$, a Humean with full access to $\xi$ could, in principle, determine $L$. However,  Humeans do not have access to the full mosaic---they are limited to macroscopic and spatiotemporal fragments of it.  
  
  As a result, Humeans faces an inverse problem:  given evidence $E$, what is the simplest law compatible with $E$ that best balances a range of epistemic guides and from which the actual mosaic can be determined?\footnote{Recall our earlier discussion of \cite{hall2009humean, hall2015humean} about the Limited Oracular Perfect Physicist (LOPP) in footnote \#9. Unlike actual Humeans, LOPP has no inverse problem to solve, as her evidence $E_{\text{LOPP}}$, combined with the assumption that it is complete, uniquely determines $(L, \xi)$. That makes her situation fundamentally different from that of actual Humeans.} Suppose the epistemic guides recommend a unique candidate law given evidence $E$:
  \begin{equation}
  	L_{\text{epistemic}}=EG(E)
  \end{equation}
where $EG(\cdot)$ is a function that maps a set of evidence to the law recommended by the epistemic guides. Taking epistemic guides seriously means having high confidence that 
  \begin{equation}\label{equal}
  	L = L_{\text{epistemic}}
  \end{equation}
However, since epistemic guides do not guarantee the correct answer, it is possible that 
 \begin{equation}\label{notequal}
  	L \neq L_{\text{epistemic}}
  \end{equation}
In probabilistic terms, a Humean who endorses PNS and  other epistemic guides should assign a high prior credence in (\ref{equal}) and a low prior credence in (\ref{notequal}). Given the high probability of (\ref{equal}), the Humean can attempt to solve an inverse problem of determining the actual mosaic, up to a point:
\begin{description}
  \item[Humean Inverse Problem] What is the actual mosaic like, given we have epistemic reasons to infer that it is optimally described by $L_{\text{epistemic}}$? 
\end{description}
This can be answered by finding:
  \begin{equation}
\xi_{\alpha} \in \Omega^{L_{\text{epistemic}}}_{\text{BSA}}, \text{with } \Omega^{L_{\text{epistemic}}}_{\text{BSA}} = \{ \xi: BS(\xi) =  L_{\text{epistemic}}  \}\footnote{In general, $\Omega^L_{\text{BSA}} \neq \Omega^L$ as some members of the latter may not be included in the former (e.g., undermining histories).}
  \end{equation}
As a final step in determining fundamental reality, Humeans infer that the actual mosaic belongs to $\Omega^{L_{\text{epistemic}}}_{\text{BSA}}$.  This rational reconstruction makes explicit how the Humean approach depends on epistemic guides. To ascertain the nature of fundamental reality---the Humean mosaic---one must collect empirical data, make ampliative inferences using epistemic guides such as PNS, and determine the likely structure of the actual mosaic based on the best available candidate for physical laws. 

Let us compare this with the rational construction on the non-Humean view of MinP.  Although MinP imposes no metaphysical restrictions on the form of fundamental laws, it is still rational to expect them to exhibit certain nice features, such as simplicity and informativeness. In BSA, these features are metaphysically constitutive of laws, whereas in MinP, they function merely as epistemic guides for discovering and evaluating candidate laws. Ultimately, these guides are defeasible---we can be fully rational in our scientific investigations and still be wrong about the fundamental laws.

The second part of \cite{chenandgoldstein}'s MinP explicitly affirms this epistemic role: 
\begin{description}
  \item[Epistemic Guides] Even though theoretical virtues such as simplicity, informativeness, fit, and degree of naturalness are not metaphysically constitutive of fundamental laws, they are good epistemic guides for discovering and evaluating them. 
\end{description}
Just as in BSA, accepting Epistemic Guides in MinP amounts to having high confidence in (\ref{equal}). A defender of MinP should be confident (though not certain) that the law recommended by epistemic guides is the governing law. At the same time, they must acknowledge the epistemic possibility that $L \neq L_{\text{epistemic}}$.  Thus, the epistemic gap in BSA is the same as that in MinP: in neither framework do we have an infallible guarantee that the laws we rationally infer are the true laws. Consequently, there is no epistemic advantage of Humeanism over non-Humeanism in terms of access to physical laws.
\begin{description}
  \item[Epistemic Parity Thesis] Humeanism does not have an epistemic advantage over non-Humeanism regarding our epistemic access to physical laws. %Humeans should not like PoI by their own lights.
\end{description}

Some Humeans might object that non-Humean views like MinP introduce additional epistemic risks \cite[p.280]{earman2005contact}, arguing that it is conceivable for us to know the entire mosaic but still remain uncertain about the laws. However, this scenario is too idealized to be relevant to real scientific practice.  A more refined objection might be that, for any given set of evidence (such as the spatiotemporally partial and macroscopic $E$), non-Humeanism allows a greater number of distinct laws than Humeanism does. However, this claim requires careful interpretation. The number of physical laws compatible with any finite body of evidence is always infinite, whether one adopts BSA or MinP. Talking about ``more'' laws in this context requires a well-defined measure. Suppose such a measure can be rigorously defined. Even then, it does not follow that the set of additional laws introduced by non-Humeanism has positive measure. A rational agent could assign epistemic probabilities such that this extra set has measure zero, yielding:
\begin{equation}
  P_{\text{BSA}}(L|E) = P_{\text{MinP}}(L|E)
\end{equation}
where $L$ is a particularly good candidate law and $E$ is our available evidence. If epistemic risks are understood probabilistically, then the Epistemic Parity Thesis remains intact. (For a similar point, but made in defense of Humeanism, see \cite{loewer2000induction}.)

The Epistemic Parity Thesis  does not rule out the possibility that non-Humeanism has an advantage over Humeanism when it comes to epistemic access to physical laws. In MinP, we assume (via PNS) that certain fundamental facts about the world are simple. In contrast, in BSA, we assume (also via PNS) that certain superficial facts (best-system laws), grounded in a complex fundamental reality, are simple. This distinction suggests that MinP makes a more plausible assumption than BSA. It is easier to believe that nature, at a deep level, is simple. It is harder to believe that nature, at a deep level, is structured in a complex way that just happens to give rise to simple best-system laws. Of course, this argument is unlikely to persuade committed Humeans, who are presumably willing to accept the consequence. However, for those who are undecided or approaching the debate for the first time, the case for non-Humeanism appears more compelling.\footnote{Thanks to Boris Kment and Tyler Hildebrand for discussion about this point. See also \cite[pp.57-58]{chenandgoldstein}. A more thorough development of this argument is left for future work.} 

Finally, this discussion connects to an on-going debate about "ratbag idealism." \cite[\S5.6]{hall2009humean} argues that, given the concern that the simplicity criterion in BSA is too subjective, Lewis and other Humeans can "perform a nifty judo move" by shifting the burden onto non-Humeans. If non-Humeans treat simplicity as an epistemic guide to laws, then, Hall suggests, they must accept that central facts of normative epistemology are also up to us. According to Hall, this is an even greater concession than the ratbag idealism of BSA. A defender of BSA might embrace ratbag idealism, treating laws as pragmatic tools for structuring our investigation of the world. From that perspective, it makes sense that what we consider to be ``laws'' is, to some degree, shaped by our pragmatic interests. However, this argument does not apply as easily to non-Humeanism. There is no compelling reason within non-Humeanism to accept that fundamental epistemological and normative facts should be "up to us" in the same way. Thus, if Hall's reasoning were correct, non-Humeans would face an even more extreme form of ratbag idealism than Humeans.

My analysis in this paper suggests that both Humeans and non-Humeans ultimately rely on strong epistemic principles such as PNS. Humeans cannot escape the problem that "central facts of normative epistemology" may be up to us unless they retreat into anti-realism about the mosaic---denying that the microscopic structure of the world, including unexplored regions of spacetime, is real. Since Humeans, too, require PNS for theory choice, they cannot execute the "nifty judo move" without undermining their own position.

\section{Conclusion}

Nomic realism is epistemically risky. There is an epistemic gap between metaphysical realism and epistemic realism. However, the gap is no smaller on Humeanism than on non-Humeanism. On both accounts, we need to decide what the physical laws are, in the vast space of possible candidates, based on our limited and macroscopic evidence about the universe. The principle of nomic simplicity serves as a fundamental epistemic guide to lawhood, directing us toward simpler laws as the most plausible candidates. Crucially, this principle is necessary for both Humean and non-Humean frameworks. It vindicates epistemic realism in cases of empirical equivalence, prevents probabilistic incoherence when dealing with nested theories, and supports realist commitments regarding induction, symmetries, dynamics, determinism, and explanation.  With many payoffs for only a small price, it is a great bargain.  

\section*{Acknowledgement} 

I am deeply grateful to Sheldon Goldstein for invaluable discussions on the nature of physical laws over the years. This paper is closely connected to several issues explored in our joint work on MinP.  For written comments and feedback, I thank Diego Arana Segura, David Builes, Nina Emery, Bosco Garcia, Noa Latham, Ned Hall, Tyler Hildebrand,  Alexander Pruss, Jonah Schupbach, Wolfgang Schwarz, Shelly Yiran Shi, Elliott Sober, and two anonymous reviewers of \textit{No\^us}. I  also appreciate helpful discussions with Emily Adlam, Jacob Barandes, James Brown, Weixin Cai, Craig Callender, Eugene Chua, Jonathan Cohen, Sam Cumming, Maaneli Derakhshani, Christopher Dorst, Frederick Eberhardt, Bill Harper, Christopher Hitchcock, Mario Hubert, Mahmoud Jalloh, Boris Kment, Barry Loewer, Kerry McKenzie, Wayne Myrvold,  Francisco Pipa, Carlo Rovelli, Daniel Rubio, Charles Sebens, Gila Sher, Chris Smeenk, Kyle Stanford, Anncy Thresher, Jennifer Wang, Kino Zhao, Shimin Zhao, participants in my graduate seminar on laws and randomness at UCSD in fall 2023, and audiences at Syracuse University, Baylor University, the University of Western Ontario, CalTech, Harvard Foundations of Physics Seminar, Simon Fraser University, and the 2023 Pacific APA for helpful discussions. This project was supported by an Academic Senate Grant from the University of California, San Diego.

%----------------------------------------------------------------------------------------
%	BIBLIOGRAPHY
%----------------------------------------------------------------------------------------

\bibliography{test}

%----------------------------------------------------------------------------------------

\end{document}